\documentstyle[prl,aps]{revtex}

\begin{document}
\title{Conditions for Nondistortion Interrogation of Quantum System}
\author{Zheng-Wei Zhou$^1$, Xingxiang Zhou$^2$, Marc J. Feldman$^2$, Guang-Can Guo$%
^1 $}
\address{$^1$Laboratory of Quantum Communication and Quantum Computation and\\
Department of Physics,\\
University of Science and Technology of China, Hefei, Anhui 230026, China\\
$^2$Superconducting Electronics Lab, Electrical and Computer Engineering\\
Department,\\
University of Rochester, Rochester, NY 14623, USA }
\maketitle

\begin{abstract}
Under some physical considerations, we present a universal formulation to
study the possibility of localizing a quantum object in a given region
without disturbing its unknown internal state. When the interaction between
the object and probe wave function takes place only once, we prove the
necessary and sufficient condition that the object's presence can be
detected in an initial state preserving way. Meanwhile, a conditioned
optimal interrogation probability is obtained.\newline
PACS numbers: 03.65.Ta, 42.50.Ct, 03.67.-a
\end{abstract}

In 1993, Elitzur and Vaidman proposed the novel concept of interaction-free
measurements(IFMs), in which the presence of an absorbing object in a
Mach-Zehnder interferometer can be inferred without apparent interaction
with the probe photon\cite{Elitzur}. As one of the counter-intuitive quantum
effects, IFMs have been extensively investigated \cite
{Kwait1,Tsegaye,Karlsson,Kwait2,Rudolph,Geszti,Jang,Mitchison1}. Aside from
their conceptual significance, IFMs are of obvious application interest too.
For instance, people applied the idea of IFM to fields such as quantum
cryptography\cite{Guo}, quantum preparation\cite{Krenn}, interaction-free
imaging\cite{White},counterfactual computation\cite{Mitchison2} and so on.

In most of these applications, it is desired that the internal state of the
object be kept unchanged. However, As emphasized by Vaidman\cite{Vaidman},
IFMs in general are not initial state preserving measurements. In the case
of a quantum object characterized by its quantum superposition, extra care
needs to be taken to avoid disturbing the initial state of the object\cite
{Potting,Xiang1,Xiang2}.

In this letter we consider a more general situation: suppose there is a
quantum ``black box'' and the task is to determine if a quantum object is in
it, without changing the (unknown) initial state of the object. We may call
such an interrogation a nondistortion interrogation of a quantum system
(abbreviated NIQS). For the purpose of NIQS, we let a quantum probe wave
partially go through the black box. If the black box is occupied with an
object a coupling interaction will entangle the originally separate systems
and make them diverge from their respective free evolutions. As a result,
the final interference pattern of the probe wave may be changed. However,
under some special conditions, some interference stripes possibly entangle
an identity transformation in the object subspace. Once the probe wave is
projected on these stripes, it will leave the object state undisturbed while
the existence of the object can be deduced. In this work we prove the
necessary and sufficient condition for the NIQS and give the conditioned
optimal probability for it, under certain physical assumptions. To obtain
our theorems the following assumptions are necessary:

a) S, the quantum system under interrogation (the quantum object in the
black box), is a metastable system whose Hilbert space is denoted as $H_S$
with dimension $n$. we assume that the state space $S(H_S)$ is closed under
free evolution.

b) The Hilbert space of the probe system $D$ , $H_D$ , is composed of two
orthogonal subspaces $H_r$ and $H_d$: $H_D=H_r\oplus H_d$. The dimension of
the space $H_d(H_r)$ is $m(m^{\prime })$. The initial probe wave function
can be written as $\left| \Psi _{probe}\right\rangle =\alpha \left| \Psi
_r\right\rangle +\beta \left| \Psi _d\right\rangle $. Here, $\left| \Psi
_r\right\rangle $ and $\left| \Psi _d\right\rangle $ are chosen from the
state space $S\left( H_r\right) $ and $S\left( H_d\right) $ respectively. We
let $|\Psi _d\rangle $ go through the black box and interact with S (if the
object is in the box), while $|\Psi _r\rangle $ is under free evolution.

c) The interaction between $D$ and $S$ is governed by a unitary operator. We
assume the time of the interaction is some known $t$. When the interaction
is over, any state driven out of the metastable state space $S(H_S)$ will
quickly decay in an irreversible way to some stable ground state $|g\rangle $
which is out of $S(H_S)$. The decay signal will be registered by some
properly arranged sensitive detectors.

In assumption b), the introduction of $H_r$ might appear to be redundant
since it does not interact with S. But it is actually vital for the purpose
of NIQS, because the interaction between $|\Psi _d\rangle $ and $|\Psi
_S\rangle $ changes the interference between $|\Psi _r\rangle $ and $|\Psi
_d\rangle $ which provides information on what is in the black box. Based on
the above assumptions, the following major steps are followed to find out if
there is an object in the black box :

i) Let the probe wave function $|\Psi _d\rangle $ go through the black box
and interact with the quantum object if it is in the box.

ii) The decay signal detectors are used to register any decay event. If a
decay is detected, we conclude that the quantum object was in the black box
but its initial state is destroyed. Otherwise, go to the next step. This is
equivalent to a partial projection measurement on the whole system $\varrho
_{tot}$: 
\begin{equation}
\varrho _{out}=P\rho _{tot}P+P_{\bot }\varrho _{tot}P_{\bot }.  \label{eq1}
\end{equation}
where the operators $P$ and $P_{\bot }$ refer to unity operators of Hilbert
space $H_S\otimes H_D$ and its complementary space $\overline{H_S\otimes H_D}
$ respectively.

iii)Perform a Von Neumann measurement on the probe wave function (in $H_D$).
Possible measurement outcomes are characterized by some projectors in an
orthogonal projector set $O$.

In step iii), we keep in mind that if nothing is in the black box the probe
will be in some definite final state corresponding to the free evolution of
the initial state. We designate the projector to that state as $P_e$. If a
successful interrogation of the object can be done, the probe will end up in
some different state. For the consideration of universality we require that
the probability of successful nondistortion interrogations of the quantum
object be independent on its (unknown) initial state.

Let the initial state of the probe wave function be $\left| \Psi
_{probe}\right\rangle =\alpha \left| \Psi _r\right\rangle +\beta \left| \Psi
_d\right\rangle $, where $\left| \Psi _r\right\rangle \in S(H_r)$ and $%
\left| \Psi _d\right\rangle \in S(H_d)$. The time evolution of the whole
system is as follows: 
\begin{equation}
\left| \Psi _{probe}\right\rangle \left| \Psi _S\right\rangle \rightarrow
\alpha e^{-i/\hbar (H^S+H^D)t}\left| \Psi _r\right\rangle \left| \Psi
_S\right\rangle +\beta e^{-i/\hbar \int_0^t(H^S+H^D+H^I)dt^{\prime }}\left|
\Psi _d\right\rangle \left| \Psi _S\right\rangle  \label{eq2}
\end{equation}
where $H^S$ and $H^D$ are the free Hamiltonian of the systems $S$ and $D$
respectively, and $H^I$ characterizes the interaction between the two
systems. When the black box is empty the interaction Hamiltonian $H^I$
vanishes and the above process reduces to the free evolution of two separate
systems. In contrast, when the black box is occupied with the quantum
object, $\left| \Psi _d\right\rangle \left| \Psi _S\right\rangle $ will be
driven by $H^I$ and a decay could happen, since any component out of the
state space $S(H_D\otimes H_S)$ will decay in an irreversible way. If no
decay signal is detected, the quantum state of the whole system is collapsed
into the following (a projection on $H_S\bigotimes H_D$): 
\begin{eqnarray}
&&\alpha e^{-i/\hbar (H^S+H^D)t}\left| \Psi _r\right\rangle \left| \Psi
_S\right\rangle +\beta (I_d\otimes I_S)e^{-i/\hbar
\int_0^t(H^S+H^D+H^I)dt^{\prime }}\left| \Psi _d\right\rangle \left| \Psi
_S\right\rangle  \nonumber \\
&=&\alpha \left| \Psi _r^{\prime }\right\rangle \left| \Psi _S^{\prime
}\right\rangle +\beta (I_d\otimes I_S)e^{-i/\hbar
\int_0^t(H^S+H^D+H^I)dt^{\prime }}e^{i/\hbar (H^S+H^D)t}(I_d\otimes
I_S)\left| \Psi _d^{\prime }\right\rangle \left| \Psi _S^{\prime
}\right\rangle  \label{eq3}
\end{eqnarray}
where $I_d$ and $I_S$ are the unity operators in the Hilbert spaces $H_d$
and $H_S$, $\left| \Psi _{r(d)}^{\prime }\right\rangle =e^{-i/\hbar
H^Dt}\left| \Psi _{r(d)}\right\rangle $ and $\left| \Psi _S^{\prime
}\right\rangle =e^{-i/\hbar H^St}\left| \Psi _S\right\rangle $. To simplify
the future calculations, the above wave function is unnormalized. In Eq(\ref
{eq3}), we see that the evolution of the system is fully specified by the
following operator $D$: 
\begin{equation}
D=(I_d\otimes I_S)e^{-i/\hbar \int_0^t(H^S+H^D+H^I)dt^{\prime }}e^{i/\hbar
(H^S+H^D)t}(I_d\otimes I_S).  \label{eq4}
\end{equation}
The following theorem provides the necessary condition for NIQS.

Theorem1: The necessary condition that an NIQS can be done is that there
exist a pair of vectors $|\chi \rangle ,\left| \Psi _d^{\prime
}\right\rangle \in S(H_d)$ which satisfy $\left\langle \chi \right| D\left|
\Psi _d^{\prime }\right\rangle =cI_S$ , where $\left| c\right| \leq 1$.

Proof: When there is no object in the box, the probe ends up in the state $%
\alpha |\Psi _r^{\prime }\rangle +\beta |\Psi _d^{\prime }\rangle $. Let $%
P_e=(\alpha |\Psi _r^{\prime }\rangle +\beta |\Psi _d^{\prime }\rangle
)(\alpha ^{*}\langle \Psi _r^{\prime }|+\beta ^{*}\langle \Psi _d^{\prime
}|)\otimes I_S$. In presence of the object, the evolution of the probe wave
function is modified. At the end of the interrogation, a measurement is done
on the probe. To make sure that a successful NIQS can be done, there must
exist a projector $P_I=|\Psi _I\rangle \langle \Psi _I|\otimes I_S$ in the
set $O$ which satisfies 
\begin{equation}
P_IP_e=0  \label{eq5}
\end{equation}
\begin{equation}
P_I(\alpha |\Psi _r^{\prime }\rangle |\Psi _S^{\prime }\rangle +\beta D|\Psi
_d^{\prime }\rangle |\Psi _S^{\prime }\rangle )=\Delta |\Psi _I\rangle |\Psi
_S^{\prime }\rangle  \label{eq6}
\end{equation}
where $\left| \Psi _I\right\rangle \in S(H_D)$ is some normalized vector of
the probe and $\Delta $ is a nonzero constant. The probability of this
outcome is $|\Delta |^2$. From Eqs. (\ref{eq5}) and (\ref{eq6}), we obtain: 
\begin{equation}
\left\langle \Psi _I\right| D\left| \Psi _d^{\prime }\right\rangle \left|
\Psi _S^{\prime }\right\rangle =(\frac \Delta \beta +\left\langle \Psi
_I|\Psi _d^{\prime }\right\rangle )\left| \Psi _S^{\prime }\right\rangle
\label{eq8}
\end{equation}
Let us introduce a wave vector $\left| \chi \right\rangle =I_d\left| \Psi
_I\right\rangle $, the projection of $|\Psi _I\rangle $ on $H_d$. Since $%
\Delta $ and $\beta $ are nonzero we deduce that $\left\| \left| \chi
\right\rangle \right\| \neq 0$. Hence, we have 
\begin{equation}
\left\langle \chi \right| D\left| \Psi _d^{\prime }\right\rangle \left| \Psi
_S^{\prime }\right\rangle =(\frac \Delta \beta +\left\langle \chi |\Psi
_d^{\prime }\right\rangle )\left| \Psi _S^{\prime }\right\rangle =c\left|
\Psi _S^{\prime }\right\rangle  \label{eq8a}
\end{equation}
where $c=\frac \Delta \beta +\left\langle \chi |\Psi _d^{\prime
}\right\rangle $. Since $\left| \Psi _S^{\prime }\right\rangle $ is
arbitrary the following must be satisfied: 
\begin{equation}
\left\langle \chi \right| D\left| \Psi _d^{\prime }\right\rangle =cI_S.
\label{eq9}
\end{equation}

In the most general case, it is a nontrivial task to determine whether two
vectors $|\chi \rangle $ and $|\Psi _d^{\prime }\rangle $ satisfying (\ref
{eq9}) exist, for a given $D$. We give some concrete discussion in the
reference\cite{Zhou}.

To find out the sufficient condition for the NIQS the operator $D$ needs to
be studied further. Under the assumption that Eq(\ref{eq9}) holds we may
define the operators $Q^{\left( i\right) }=Tr_S\left[ D\left| \Psi
_d^{\prime }\right\rangle \left| i\right\rangle \left\langle i\right|
\left\langle \Psi _d^{\prime }\right| D^{+}\right] $. Here, $\{\left|
i\right\rangle ,i=1,...,n\}$ is a set of orthogonal bases in the Hilbert
space $H_S$. In the Hilbert space $H_d$, the kernel space of the operator $%
Q^{\left( i\right) }$ is denoted as $K_i$. The intersection of all the $n$
kernel spaces is $K=\bigcap_{i=1}^nK_i$. We denote the dimension of the
space $\overline{K}$, which is the complementary space of $K$ in $H_d$, by $%
l(l\leq m)$. We pick up some set of orthonormal states $\{\left| \chi
\right\rangle ,\left| \chi _1\right\rangle ,...,\left| \chi
_{l-1}\right\rangle \}$ spanning the space $\overline{K}$. Then, Eq(\ref{eq9}%
) can be expressed in the following alternative way: 
\begin{equation}
D\left| \Psi _d^{\prime }\right\rangle \left| \Psi _S^{\prime }\right\rangle
=c\left| \chi \right\rangle \left| \Psi _S^{\prime }\right\rangle
+\sum_{j=1}^{l-1}\left| \chi _j\right\rangle \left| m_{S(j)}\right\rangle .
\label{eq10}
\end{equation}
Here, $\left| m_{S(j)}\right\rangle =\left\langle \chi _j\right| D\left|
\Psi _d^{\prime }\right\rangle \left| \Psi _S^{\prime }\right\rangle $. By
making use of Eq(\ref{eq10}) we may provide our main result.

Theorem 2: The necessary and sufficient condition for the NIQS is that Eq(%
\ref{eq10}) holds and $\left| \Psi _d^{\prime }\right\rangle -c\left| \chi
\right\rangle $ is linearly independent of the state set $\{\left| \chi
_j\right\rangle ;j=1,...,l-1\}$.

Proof: We have proved that Eq(\ref{eq10}) is the necessary condition. If Eq(%
\ref{eq10}) holds, in Hilbert space $H_S\otimes H_D$ the final state of the
whole system is: 
\begin{equation}
\left| \Psi _{probe}\right\rangle \left| \Psi _S\right\rangle \rightarrow
\alpha \left| \Psi _r^{\prime }\right\rangle \left| \Psi _S^{\prime
}\right\rangle +\beta c\left| \chi \right\rangle \left| \Psi _S^{\prime
}\right\rangle +\beta \sum_{j=1}^{l-1}\left| \chi _j\right\rangle \left|
m_{S(j)}\right\rangle .  \label{eq11}
\end{equation}
Considering Eqs. (\ref{eq5}) and (\ref{eq6}), we have: 
\begin{equation}
\left\langle \Psi _I\right| \left( \alpha \left| \Psi _r^{\prime
}\right\rangle +\beta \left| \Psi _d^{\prime }\right\rangle \right) =0
\label{eq12}
\end{equation}
\begin{equation}
\left\langle \Psi _I|\chi _j\right\rangle =0  \label{eq13}
\end{equation}
\begin{equation}
\left\langle \Psi _I\right| \left( \alpha \left| \Psi _r^{\prime
}\right\rangle +c\beta \left| \chi \right\rangle \right) =\Delta .
\label{eq14}
\end{equation}
Subtracting Eq(\ref{eq14}) from Eq(\ref{eq12}) we obtain 
\begin{equation}
\left\langle \Psi _I\right| \beta \left( \left| \Psi _d^{\prime
}\right\rangle -c\left| \chi \right\rangle \right) =-\Delta .  \label{eq15}
\end{equation}
$\Delta \neq 0$ requires that $\left| \Psi _d^{\prime }\right\rangle
-c\left| \chi \right\rangle $ be linearly independent of the set of vectors $%
\{\left| \chi _j\right\rangle ;j=1,...,l-1\}$.

Now if Eq(\ref{eq10}) holds and $|\Psi _d^{\prime }\rangle -c\left| \chi
\right\rangle $ is linearly independent of the state set $\{\left| \chi
_j\right\rangle ;j=1,...,l-1\}$, we may prove the sufficient condition by
constructing a projector $P_I$ satisfying Eq(\ref{eq5}) and Eq(\ref{eq6}).
We do this by using the Schmidt orthogonalization process. First we define
the state set $N$ consisting of $l+1$ normalized vectors $\{\alpha \left|
\Psi _r^{\prime }\right\rangle +\beta \left| \Psi _d^{\prime }\right\rangle
,\gamma \left( \alpha \left| \Psi _r^{\prime }\right\rangle +c\beta \left|
\chi \right\rangle \right) ,\left| \chi _j\right\rangle ;j=1,...,l-1\}$,
where $\gamma =\frac 1{\left\| \alpha \left| \Psi _r^{\prime }\right\rangle
+c\beta \left| \chi \right\rangle \right\| }$ is the normalization
coefficient for $\alpha \left| \Psi _r^{\prime }\right\rangle +c\beta \left|
\chi \right\rangle $. We assume that $\alpha \neq 0$. Since $\left| \chi
_1\right\rangle ,...\left| \chi _{l-1}\right\rangle $ and $\left| \chi
\right\rangle $ are orthogonal to each other and $\left| \Psi _d^{\prime
}\right\rangle -c\left| \chi \right\rangle $ is linearly independent of $%
\{\left| \chi _j\right\rangle ;j=1,...,l-1\}$, we may deduce that all
vectors in the state set $N$ are linearly independent. To make an
orthonormal set out of $N$, we let the first $l-1$ vectors be $\{\left| \chi
_j\right\rangle ;j=1,...,l-1\}$. We then calculate the $l$th vector using $%
\alpha \left| \Psi _r^{\prime }\right\rangle +\beta \left| \Psi _d^{\prime
}\right\rangle $: $\left| \widetilde{\Psi }\right\rangle =\gamma ^{\prime
}\left( \alpha |\Psi _r^{\prime }\rangle +\beta |\Psi _d^{\prime }\rangle
-\Sigma _{i=1}^{l-1}\left\langle \chi _i\right| (\alpha |\Psi _r^{\prime
}\rangle +\beta \Psi _d^{\prime }\rangle )\left| \chi _i\right\rangle
\right) $, where $\gamma ^{\prime }$ is the normalization coefficient.
Similarly, we can obtain the ($l+1$)th vector $\left| \Psi _I\right\rangle
=\gamma ^{\prime \prime }\left( \gamma \left( \alpha \left| \Psi _r^{\prime
}\right\rangle +c\beta \left| \chi \right\rangle \right) -\left\langle 
\widetilde{\Psi }\right| \gamma \left( \alpha \left| \Psi _r^{\prime
}\right\rangle +c\beta \left| \chi \right\rangle \right) \left| \widetilde{%
\Psi }\right\rangle \right) $. It is then obvious that the projector $%
P_I=|\Psi _I\rangle \langle \Psi _I|\otimes I_S$ satisfies Eqs(\ref{eq5})
and (\ref{eq6}). We thus complete our proof.

To clarify the connotation of the above theorems let us consider the
following physical picture of NIQS: the quantum object in the black box can
be seen as a ``scattering object'' corresponding to the probe wave. Due to
the interaction with the object the probe wave will change its initial
coherence. On the other hand, the quantum object is also affected by the
probe wave. Thus, each scattering wave component entangles different
evolution of the object. If we know all the information on the evolution of
the composite system it possibly allows us to choose a proper probe wave
such that a successful scattering wave component can be produced. For the
purpose of NIQS this component should correspond to the free evolution of
the object, at the same time, be orthogonal to any other scattering wave
components. ( The form of this component can be obtained by using Schmidt
orthogonalization steps outlined in the proof of theorem 2.) Once this
scattering wave component is registered by a detector we may deduce the
existence of the object in the internal preserving way. In fact, theroem 2
provides the necessary and sufficient condition for the existence of this
scattering wave component.

Let us denote the Hilbert space spanned by $\left\{ |\Psi _r^{\prime
}\rangle ,|\chi \rangle ,|\chi _1\rangle ,...|\chi _{l-1}\rangle \right\} $
as $H_{l+1}$ and its complementary space in $H_D$ as $H_{l+1}^{\bot }$. For
the success probability of an NIQS we have the following corollary:

Corollary 1: $P_I$ obtained by using Schmidt orthogonalization process in $%
H_{l+1}$ (as outlined above) maximizes the success probability $\left|
\Delta \right| ^2$ for a given $\alpha $.

Proof: Let $I_{l+1}$ and $I_{l+1}^{\bot }$ be the unity operators of $%
H_{l+1} $ and $H_{l+1}^{\bot }$ respectively. Any Von Neumann projector
characterizing a successful nondistortion interrogation output can be
written as $P_I^{\prime }=|\Psi _I^{\prime }\rangle \langle \Psi _I^{\prime
}|\otimes I_S$ ($P_I^{\prime }$ satisfies (5), (6)). Furthermore, we have
the following relation: 
\begin{eqnarray}
P_I^{\prime } &=&(I_{l+1}\otimes I_S)P_I^{\prime }(I_{l+1}\otimes
I_S)+(I_{l+1}^{\bot }\otimes I_S)P_I^{\prime }(I_{l+1}^{\bot }\otimes I_S) 
\nonumber  \label{eq16} \\
\ &=&dP_I+(I_{l+1}^{\bot }\otimes I_S)P_I^{\prime }(I_{l+1}^{\bot }\otimes
I_S)  \label{eq16}
\end{eqnarray}
where $|d|\leq 1$. The second equality is because the projector $P_I$
satisfying the NIQS conditions ( Eqs(\ref{eq12})-(\ref{eq14})) in $%
H_{l+1}\otimes H_S$ is unique. On the other hand, the second term in the
last line will never contribute to the projection probability when $%
P_I^{\prime }$ operates on $\alpha |\Psi _r^{\prime }\rangle |\Psi
_S^{\prime }\rangle +\beta D|\Psi _d^{\prime }\rangle |\Psi _S^{\prime
}\rangle $. Hence, the success probability of $P_I^{\prime }$ is: 
\begin{eqnarray}
\Pr ob(\alpha )_{P_I^{\prime }} &=&\left| \left\langle \Psi _I^{\prime
}\right| (\alpha \left| \Psi _r^{\prime }\right\rangle +c\beta \left| \chi
\right\rangle )\right| ^2  \nonumber \\
\ &=&|d|^2\left| \left\langle \Psi _I\right| (\alpha \left| \Psi _r^{\prime
}\right\rangle +c\beta \left| \chi \right\rangle )\right| ^2\leq \Pr
ob(\alpha )_{P_I}  \label{eq17}
\end{eqnarray}

The above corollary shows that the maximal success probability for the NIQS
process is just the probability of this successful scattering wave component
when the original probe wave is given. We note that the vectors $\left| \Psi
_d^{\prime }\right\rangle $ and $\left| \chi \right\rangle $ satisfying Eq(%
\ref{eq9}) may not be unique. However, once $\left| \chi \right\rangle $ and 
$\left| \Psi _d^{\prime }\right\rangle $ are chosen we may obtain the
optimal success probability for these two vectors, by Schmidt
orthogonalization steps .

Corollary 2: Under the condition that the wave function $\left| \Psi
_d^{\prime }\right\rangle $ and $\left| \chi \right\rangle $ are given the
optimal success probability of the NIQS is as follows: 
\begin{equation}
P_{opt}=\max_{\left| \alpha \right| \in \left[ 0,1\right) }\Pr ob(\alpha
)_{P_I}.  \label{eq18}
\end{equation}

To elucidate our result in a more concrete way, let us consider an example
recently investigated \cite{Potting,Xiang1,Xiang2}. As in Fig. 1, the atom
is prepared in an arbitrary superposition of the two metastable states $%
|m_{+}\rangle $ and $|m_{-}\rangle $. By absorbing a $+$ or $-$ (circularly)
polarized photon the atom can make a transition to the excited state $%
|e\rangle $. It then decays rapidly to the ground state $|g\rangle $ in an
irreversible way. Now, without disturbing the initial state of the atom we
want to find out if it is in the black box located in the lower arm of the
Mach-Zehnder interferometer. To do that, we use an $x$ polarized photon $%
\frac 1{\sqrt{2}}(\left| l_{+}\right\rangle +\left| l_{-}\right\rangle )$ as
the probe and direct it into the interferometer. The amplitude transmission
and reflection coefficients of the two beam splitters are $(\alpha ,\beta )$
and $(\beta ,\alpha )$ respectively so that the photon exits from the upper
port of $PBS_2$ with certainty when the atom is not in the black box. $PBS_1$
changes the state of the incident photon to $\left| \Psi
_{probe}\right\rangle =\frac \alpha {\sqrt{2}}\left( \left|
u_{+}\right\rangle +\left| u_{-}\right\rangle \right) +\frac \beta {\sqrt{2}}%
\left( \left| l_{+}\right\rangle +\left| l_{-}\right\rangle \right) $, where 
$l(u)$ refers to the lower (upper) optical path and $+(-)$ refers to the
polarization of the photon. After the first beam splitter, the photon wave
function is in a superposition of the {\em u}pper and {\em l}ower branches.
These correspond to $H_r$ and $H_d$ in our formulation. Here $H_r,H_d$ are
two dimensional subspaces with base vectors $|u_{\pm }\rangle $ and $\left|
l_{\pm }\right\rangle $. $H_S$, the metastable state space of the atom, is a
two dimensional space spanned by $\left| m_{+}\right\rangle $ and $\left|
m_{-}\right\rangle $. The interaction between $H_d$ and $H_S$ and the
afterward dissipation is characterized by the operator $D=\left|
l_{+}\right\rangle \left| m_{-}\right\rangle \left\langle m_{-}\right|
\left\langle l_{+}\right| +\left| l_{-}\right\rangle \left|
m_{+}\right\rangle \left\langle m_{+}\right| \left\langle l_{-}\right| $. We
may further deduce that the space $\overline{K}$ is just $H_d$. Thus, when
choosing $\left| \Psi _d^{\prime }\right\rangle =\left| \chi \right\rangle =%
\frac 1{\sqrt{2}}(\left| l_{+}\right\rangle +\left| l_{-}\right\rangle )$
and $\left| \chi _1\right\rangle =\frac 1{\sqrt{2}}(\left|
l_{+}\right\rangle -\left| l_{-}\right\rangle )$ we find $\left\langle \chi
\left| D\right| \Psi _d^{\prime }\right\rangle =\frac 12\left( \left|
m_{-}\right\rangle \left\langle m_{-}\right| +\left| m_{+}\right\rangle
\left\langle m_{+}\right| \right) =\frac 12I_S$ and $\left| \Psi _d^{\prime
}\right\rangle -\frac 12\left| \chi \right\rangle $ is linearly independent
of the state vector $\left| \chi _1\right\rangle $. Therefore, the
sufficient and necessary condition for NIQS is satisfied. Following the
Schmidt orthogonalization process we can obtain the projector $P_I$: $%
P_I=\left( \frac{\beta ^{*}}{\sqrt{2}}\left( \left| u_{+}\right\rangle
+\left| u_{-}\right\rangle \right) -\frac{\alpha ^{*}}{\sqrt{2}}\left(
\left| l_{+}\right\rangle +\left| l_{-}\right\rangle \right) \right) \left( 
\frac \beta {\sqrt{2}}\left( \left\langle u_{+}\right| +\left\langle
u_{-}\right| \right) -\frac \alpha {\sqrt{2}}\left( \left\langle
l_{+}\right| +\left\langle l_{-}\right| \right) \right) \otimes I_S$. Here,
the actual measurement of the probe photon is performed after $PBS_2$ with
the projector $P_I^{\prime }=UP_IU^{\dag }=\frac 12\left( \left( \left|
l_{+}\right\rangle +\left| l_{-}\right\rangle \right) \left( \left\langle
l_{+}\right| +\left\langle l_{-}\right| \right) \right) \otimes I_S$, where
the unitary operator U describes the effect of $PBS_2$ on the probe photon.
Thus, the success probability is $\Pr ob(\alpha )=\frac 14\left| \alpha
\right| ^2\left| \beta \right| ^2$. When $\left| \alpha \right| =\left|
\beta \right| =\frac 1{\sqrt{2}}$ the optimal probing probability is
obtained: $P_{opt}=\frac 1{16}$. So, when the lower detector fires we can
deduce that the atom is in the black box, with success probability 1/16.
This example has just been investigated by P\"otting et. al \cite{Potting}.

\begin{figure}[tbp]
\caption{A nondistortion interrogation of an atom prepared in an arbitrary
superposition of the metastable states $|m_{+}\rangle $ and $|m_{-}\rangle $%
, which are coupled to the excited state $|e\rangle $ through $+$ or $-$
(circularly) polarized photons. An $x$ polarized photon is used as the
probe. The black box is located in the lower optical path. }
\label{fig:figure1}
\end{figure}

In our scheme we assume that the interaction between the object and probe
system occurs only once. It has been shown that under certain conditions a
higher efficiency can be obtained in an iterative way\cite
{Kwait1,Tsegaye,Kwait2,Jang,Xiang2}. Some results on this particular scheme
have been obtained in ref \cite{Zhou}.

It should be emphasized that the significance of NIQS lies in the coherence
of the detected quantum state being preserved in the Hilbert space $H_S$. In
other words, quantum information in Hilbert space $H_S$ will not be
contaminated by the probing process. This way to manipulate quantum objects
is of potential application in the recently developed quantum information
science. Since the detected system need not be restricted to pure states we
may also cast our interests on mixed states. As pointed out by P\"otting et.
al \cite{Potting} the nondistortion interrogation provides a tool to monitor
a subsystem in a many-particle system without destroying the entanglement
between the particles.

In this letter we have studied the process of NIQS under some physical
assumptions. We proved the necessary and sufficient condition for NIQS
process in our formulation. We obtained the optimal success probability of
NIQS when the state vectors $\left| \Psi _d^{\prime }\right\rangle $ and $%
\left| \chi \right\rangle $ are given. We also showed that our results apply
to IFMs which is a special case of the problem we discussed. As a novel
method to manipulate quantum systems NIQS may be applied in future quantum
information processing.

\vskip 5mm

Work of Z.W.Z. and G.C.G. was funded by National Natural Science Foundation
of China, National Fundamental Research Program, and also by the outstanding
Ph. D thesis award and the CAS's talented scientist award entitled to Luming
Duan. Research of X.Z. and M.J.F. was supported in part by ARO grant
DAAG55-98-1-0367.

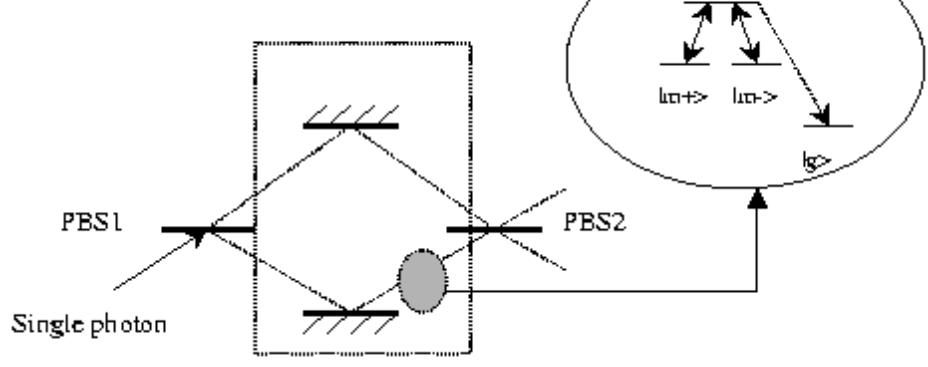
\end{document}